\documentclass[10pt,conference,a4paper]{IEEEtran}
\usepackage{times,amsmath,epsfig}
\usepackage{epstopdf}
\usepackage[]{graphicx}
\usepackage{amsmath,amssymb}
\usepackage{subfig}
\usepackage{caption}
\usepackage[utf8]{inputenc}      %
\usepackage[T1]{fontenc}
\usepackage[english]{babel}       %
\usepackage{esvect}       %
\usepackage{blindtext}
\usepackage{subscript}
\usepackage{nicefrac}
\usepackage{units}
\usepackage{setspace}           %
\usepackage{paralist} %
\usepackage[export]{adjustbox}   %
\usepackage{multicol}
\usepackage{xspace}               %
\usepackage{relsize}    %
\def\sc#1{%
  \smaller%
  #1%
  \larger\xspace%
}
\usepackage{tikz}   %
\usetikzlibrary{shapes,arrows,shadows,positioning,backgrounds}

\graphicspath{{./fig/}}

\newcommand{\eg}{\emph{e.\,g.}\xspace}

\newcommand{\ie}{\emph{i.\,e.}\xspace}
\newcommand{\mcc}[1]{\multicolumn{1}{c}{#1}} %
\newcommand{\mccl}[1]{\multicolumn{1}{|c}{#1}} %
\newcommand{\mccr}[1]{\multicolumn{1}{c|}{#1}} %

\newcommand{\roicontrolrate}{\unitfrac[1--3]{Mbit}{s}\xspace}

\newcommand{\hevcskiprate}{\unitfrac[0.7--1.0]{Mbit}{s}\xspace}  %

\newcommand{\hevcskipimprovementmeaninteger}{\unit[32]{\%}\xspace}

\newcommand{\hevcskipimprovementbigblocksmeaninteger}{\unit[38]{\%}\xspace}

\newcommand{\tabref}[1]{Table~\ref{#1}}

\newcommand{\figref}[1]{Fig.~\ref{#1}}

\newcommand{\secref}[1]{Section~\ref{#1}}

\newcommand{\Secref}[1]{Section~\ref{#1}}

\newcommand{\equref}[1]{Equation~\eqref{#1}}

\title{Region of Interest Coding for Aerial \\ Surveillance Video Using AVC \& HEVC}

\author{%
{Holger Meuel, Florian Kluger and Jörn Ostermann}%
\vspace{1.6mm}\\
\fontsize{10}{10}\selectfont\itshape
Institut für Informationsverarbeitung \\Gottfried Wilhelm Leibniz Universität Hannover,
Germany\\
\fontsize{9}{9}\selectfont\ttfamily\upshape
\{meuel, kluger, office\}@tnt.uni-hannover.de\\
\vspace{1.2mm}\\
}

\begin{document}
\maketitle

%
%

\begin{abstract}
Aerial surveillance from \emph{Unmanned Aerial Vehicles} (\sc{UAV}{}s), \ie with moving cameras, is of growing interest for police as well as disaster area monitoring.
For more detailed ground images the camera resolutions are steadily increasing. 
Simultaneously the amount of video data to transmit is increasing significantly, too.
To reduce the amount of data, \emph{Region of Interest} (\sc{ROI}) coding systems were introduced which mainly encode some regions in higher quality at the cost of the remaining image regions.
We employ an existing \sc{ROI} coding system relying on \emph{global motion compensation} to retain full image resolution over the entire image.
Different \sc{ROI} detectors are used to automatically classify a video image on board of the \sc{UAV} in \sc{ROI} and non-\sc{ROI}.
We propose to replace the modified \emph{Advanced Video Coding} (\sc{AVC}) video encoder by a modified \emph{High Efficiency Video Coding} (\sc{HEVC}) encoder.
Without any change of the detection system itself, but by replacing the video coding back-end we are able to improve the coding efficiency by \hevcskipimprovementmeaninteger{} on average although regular \sc{HEVC} provides coding gains of \unit[12--30]{\%} only for the same test sequences and similar \sc{PSNR} compared to regular \sc{AVC} coding.
Since the employed \sc{ROI} coding mainly relies on intra mode coding of new emerging image areas, gains  of \sc{HEVC-ROI} coding over \sc{AVC-ROI} coding compared to regular coding of the entire frames including predictive modes (\emph{inter}) depend on sequence characteristics.
We present a detailed analysis of bit distribution within the frames to explain the gains.
In total we can provide coding data rates of \hevcskiprate for full \sc{HDTV} video sequences at \unit[30]{fps} at reasonable quality of more than \unit[37]{dB}.
\\[1\baselineskip] %
\end{abstract}

\begin{IEEEkeywords}
  Region of Interest (ROI) Video Coding, HEVC, Global Motion Compensation (GMC), Moving Object Detection, UAV Attached Moving Camera, Aerial Surveillance
\end{IEEEkeywords}

\section{Introduction}
\label{sec:introduction}

\fboxsep0pt
\fboxrule0.5pt
\begin{figure}[t]
  \centering
  \subfloat[Detection of new areas by global motion compensation (\sc{GMC}).\label{sfig:new_area_detection}]
  {
    \hspace{5pt} \scalebox{0.9}{ \includegraphics[width=\linewidth]{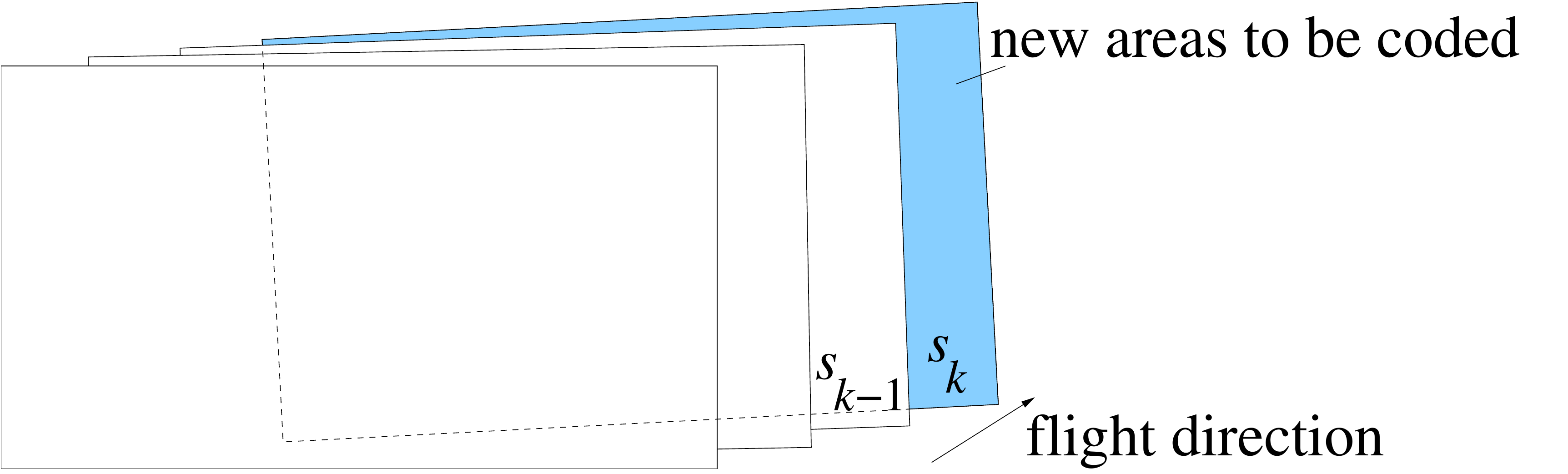} }
  }
  \\
  \subfloat[Detection of moving objects (diff. image-based).\label{sfig:mo_detection}]
  {
    \includegraphics[width=0.20\textwidth]{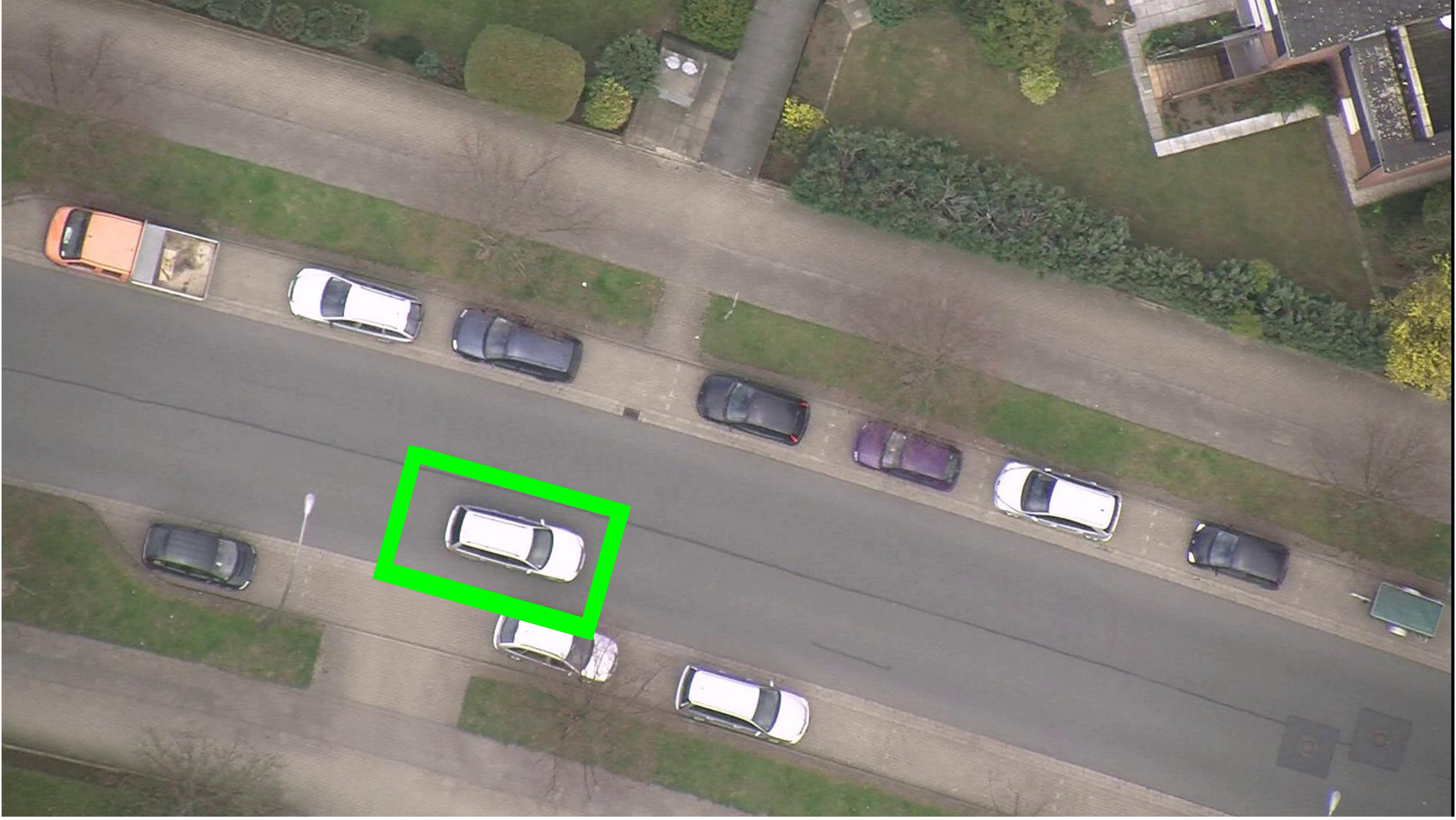}
  }
  \quad
  \subfloat[Transmission of macro- blocks/\sc{CTU}{}s containing \sc{ROI}. \label{sfig:roi_to_transmit}]
  {
    \fbox{\includegraphics[width=0.20\textwidth]{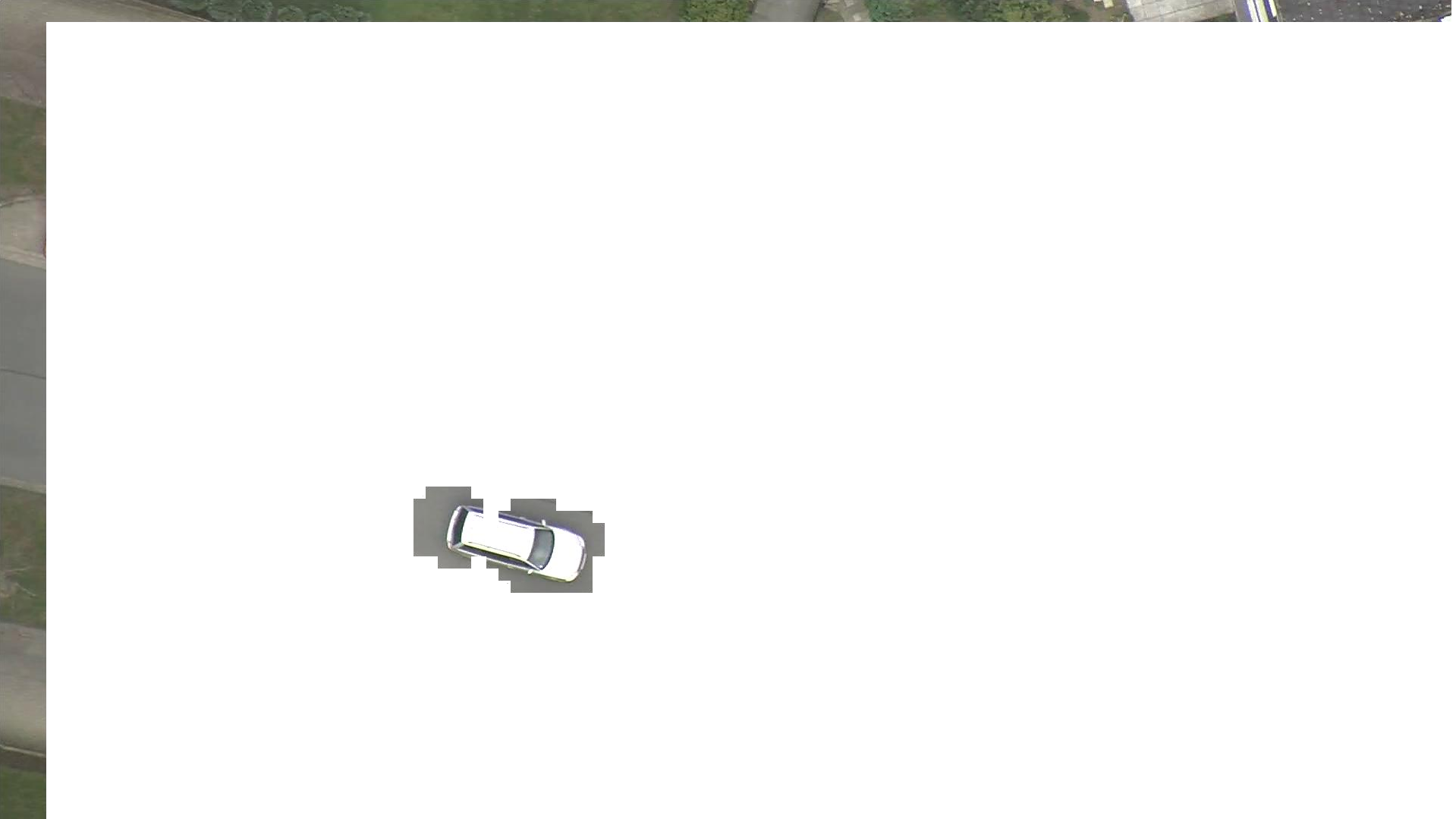}}
  }
  \caption{
    Illustration of \sc{ROI} detection and coding.
    \vspace*{-1.75\baselineskip}
    \label{fig:illustrationROIsystem}}
\end{figure}

In aerial surveillance applications from \emph{Unmanned Aerial Vehicles} (\sc{UAV}{}s) a small encoded video data rate is as important as a high quality and resolution of the observed area.
\emph{Region of Interest} (\sc{ROI}) coding is a common solution for reducing the coding bit rate at the cost of certain image areas which are considered to be less important (\ie the background, non-\sc{ROI}) than others (\ie the foreground, \sc{ROI}) \cite{roi_coding_buchkapitel_shortref_short}.
One challenge in an \sc{UAV} mounted system is to classify \sc{ROI} and non-\sc{ROI} fully automatically in order to assign quality levels and bit rates for different image areas.
Finally, a coding scheme is needed which allows to assign different quality levels within one frame.
The video coding system in \cite{roi_coding_meuel_munderloh} avoids distinguishing different quality levels retaining full \sc{HDTV} ground resolution at a data rate of \roicontrolrate.
This coding system relies on global motion compensation of the background and encoding and transmission of \emph{New Areas} (\sc{NA}{}s) contained in the current frame but not in the previous one.
To retain also \emph{local} movement in the decoded video, \emph{Moving Objects} (\sc{MO}{}s) are encoded and transmitted additionally.
Those two types of \sc{ROI}{}s are automatically detected by special \sc{ROI} detectors, one for \sc{NA}{}s and one for \sc{MO}{}s.
By the modular design of this system it is possible to include additional \sc{ROI} detectors like shape based detectors \cite{target_recognition_from_remote_sensing_images} or replace the \sc{ROI-MO} detector \eg with a motion vector based \sc{MO} detector \cite{mo_detection_and_tracking_by_spatio_temporal_motion_vector_graphs_short}. %
The video coder itself consists of an externally controlled \emph{Advanced Video Coding} (\sc{AVC} \cite{avc_short}) \emph{x264} encoder, which sets any non-\sc{ROI} area to \emph{skip mode} and thus introduces no extra transmission cost by preserving standard compliance for the bit stream.
Alternative \sc{ROI} coding systems propose the variation of the \sc{QP} for \sc{ROI}/non-\sc{ROI} areas on a macroblock/\emph{Coding Unit} (\sc{CU}) level, respectively \cite{roi_based_resource_allocation_for_conversational_video_compression_of_H264_AVC}, which unintentionally introduces lots of extra transmission cost for signaling of the \sc{QP} changes for numerous non-connected \sc{ROI}{}s \cite{scc_meuel_schmidt}. %
Other \sc{ROI} coding schemes replace the \emph{Rate-Distortion Optimization} within the \sc{HEVC} encoder in order to assign a different amount of bits to \sc{ROI} and non-\sc{ROI} \cite{perceptual_conversational_roi_video_coding, region_of_interest_based_conversational_hevc_coding_with_hieararchical_perception_model_of_face_short}. %
However, when employing a global motion compensation postprocessing, all data from non-\sc{ROI} area is discarded anyway at the decoder and thus the optimal bit allocation scheme obviously is to spend as much bits as possible on \sc{ROI} and as few bits as possible on non-\sc{ROI} areas. %
This constraint can best be fulfilled by employing the skip mode like in the reference system \cite{roi_coding_meuel_munderloh} why we decided for a skip-implementation in the \sc{HEVC} reference software \sc{HM 10.0} similar to the \sc{AVC}-based coding back-end. %

In this paper we propose the replacement of the video encoder by an externally controlled \emph{High Efficiency Video Coding} (\sc{HEVC} \cite{hevc_v1_standard}) encoder \cite{mesh_roi-coding_journal_pcs2013_followup_meuel_munderl_reso}. %
We demonstrate an efficient mode control including the \emph{skip mode} and the mandatory \sc{HEVC} syntax elements \emph{merge flag} and \emph{merge index}.
Moreover we present a detailed analysis of the spatial bit distribution.

The remaining paper is organized as follows:
\Secref{sec:RoiCodingReferenceSystem} summarizes the \sc{ROI} coding system shortly, and explains the encoding process in detail.
The proposed replacement of the coding back-end towards \sc{HEVC} and implementation details are given in \secref{ssec:hevcCodingBackendImplementation}. 
Experimental results are discussed in \secref{sec:experiments} before \secref{sec:conclusion} concludes the paper.

\section{ROI-based Reference Coding System}
\label{sec:RoiCodingReferenceSystem}

\begin{figure}
  \begin{center}
    \hspace*{-0.25cm}
    \scalebox{0.945}{ \includegraphics[width=\linewidth]{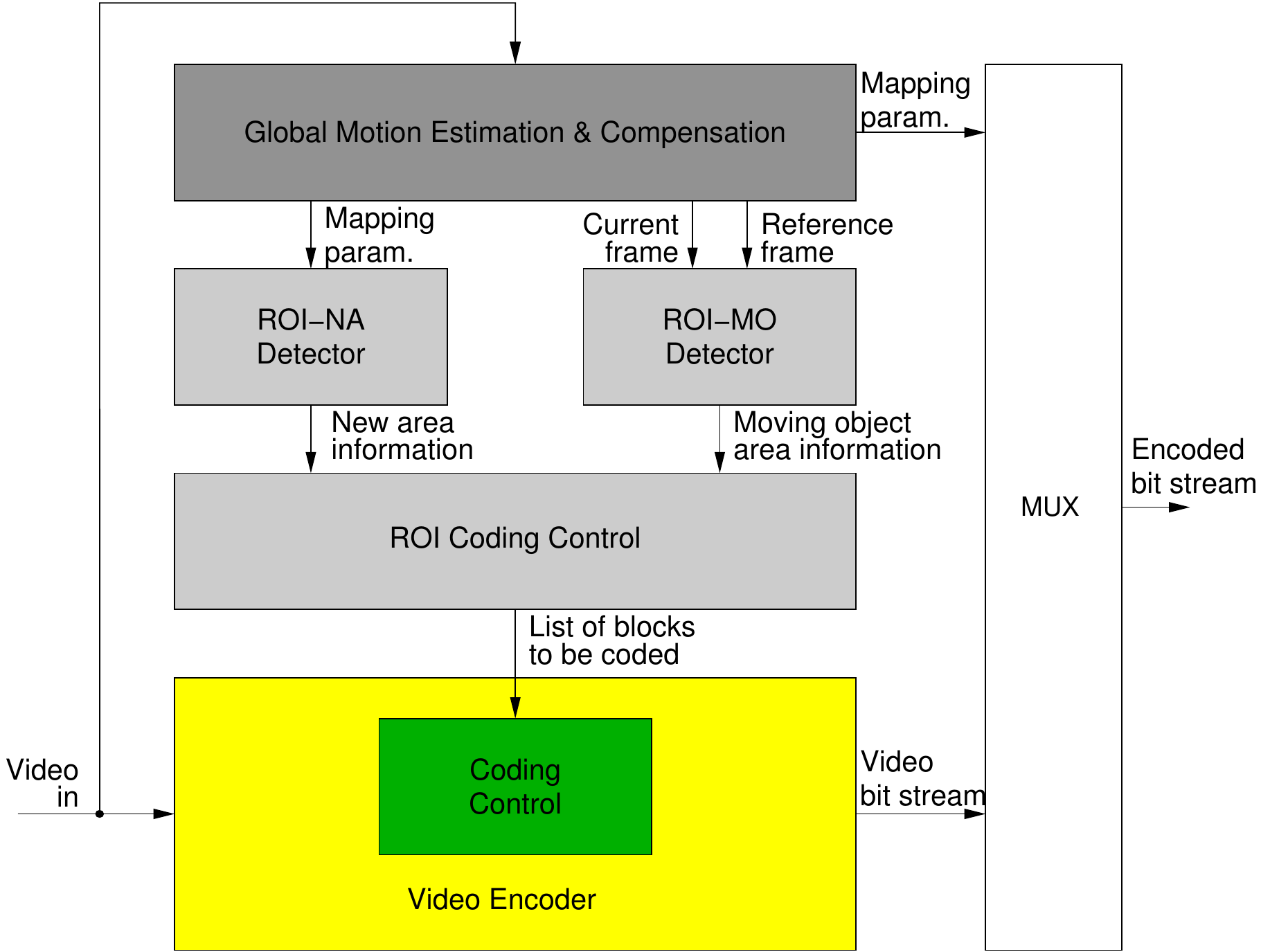} } %
  \end{center}
  \caption{Block diagram of \sc{GME/GMC}-based \sc{ROI} coding system. Gray: unmodified (dark: \sc{GME/GMC}, light: \sc{ROI} detection), yellow/green: external controlled video encoder (based on \cite{roi_coding_meuel_munderloh}).
\vspace*{-1.75\baselineskip}
  }
  \label{fig:roi-control-scheme}
\end{figure}

Although the \sc{ROI} detection system remains unchanged compared to the reference \cite{roi_coding_meuel_munderloh} like afore-mentioned we summarize the system before we focus on the (\sc{AVC}-based) coding back-end and the proposed upgrade to \sc{HEVC} in \secref{ssec:hevcCodingBackendImplementation}.

The idea of data reduction is to exploit the special characteristic of a \emph{planar} landscape of aerial surveillance videos which is true for high flight altitudes  (\figref{fig:illustrationROIsystem}).
Assuming a planar landscape, one frame $k-1$ can be projected into the consecutive frame $k$ employing a projective transform with 8 parameters $\vv*{a}{k} = ( a_{1,k}, a_{2,k}, \dots, a_{8,k})^T$. %
The pixel coordinates from the preceding frame $\vv*{p}{k-1} = (x_{k-1}, y_{k-1})$ are mapped to the position $\vv*{p}{k} = (x_{k},y_{k})$ of the current one with the mapping parameter set $\vv*{a}{k}$ \eqref{eq:projective_trafo}.

\begin{footnotesize}

\begin{align}
\label{eq:projective_trafo}
\hspace*{-0.95em}
\textstyle
F \left(\vv{p},\vv*{a}{k} \right) = \frac{a_{1,k} \cdot x_{k-1} + a_{2,k} \cdot y_{k-1} + a_{3,k}}{a_{7,k} \cdot x_{k-1} + a_{8,k} \cdot y_{k-1} + 1},
\textstyle
\frac{a_{4,k} \cdot x_{k-1} + a_{5,k} \cdot y_{k-1} + a_{6,k}}{a_{7,k} \cdot x_{k-1} + a_{8,k} \cdot y_{k-1} + 1} 
\end{align}
\end{footnotesize}

To determine $\vv*{a}{k}$, first, a global motion estimation is performed.
To do so, \emph{Harris Corners} \cite{harris_corner_detector_short} are used to defined a set of good-to-track feature points in the frame $k$. %
A \emph{Kanade-Lucas-Tomasi} (\sc{KLT}) \cite{tomasi_kanade_detection_and_tracking_of_point_features, shi_tomasi_good_features_to_track} feature tracker is employed afterwards to relocate the feature positions in the frame $k-1$ and thereby generate a sparse optical flow between the frames. %
Outliers such as false tracks are removed and the final mapping parameter set $\vv*{a}{k}$ is determined by \emph{Random Sample Consensus} (\sc{RANSAC}) \cite{fischler_bolles-1981_short}. 
This mapping parameter set is used for the \emph{Global Motion Compensation} (\sc{GMC}) as the first block in the block diagram of the coding system (\figref{fig:roi-control-scheme}) by employing \equref{eq:projective_trafo}.
The mapping parameter set $\vv*{a}{k}$ is further employed to determine the \emph{New Area} (\sc{NA}) in the current image $k$ by the \emph{\sc{ROI-NA} Detector}. %
In order to detect local motion, the pel-wise difference image between the global motion compensated frame $\hat{k}$ and the current frame $k$ is calculated and spots of high energy are considered to be moving objects (\figref{sfig:mo_detection}). %
Both \sc{ROI}{}s are passed to the \emph{\sc{ROI} Coding Control} block which basically assigns the pel-wise \sc{ROI} to the corresponding macroblocks for \sc{AVC} coding (\figref{sfig:roi_to_transmit}).
Any \sc{ROI} macroblock is \sc{AVC} encoded as usual whereas any non-\sc{ROI} macroblock is forced to \emph{skip mode}.
Thus, the data rate is significantly reduced while standard compliance of the bit stream is retained.
The mapping parameter set has to be transmitted in the data stream as well which could be realized by encapsulating the 8 parameters per frame in \emph{Supplemental Enhancement Information} (\sc{SEI}) messages.
However, after decoding of the bit stream a postprocessing is necessary in order to align \sc{ROI}{}s from the current frame within the reconstructed background from the previous frames \cite{roi_coding_buchkapitel_shortref_short}. %

\begin{figure}
  \centering
  \hspace*{-0.5cm}
  \scalebox{1}{ \includegraphics[width=\linewidth]{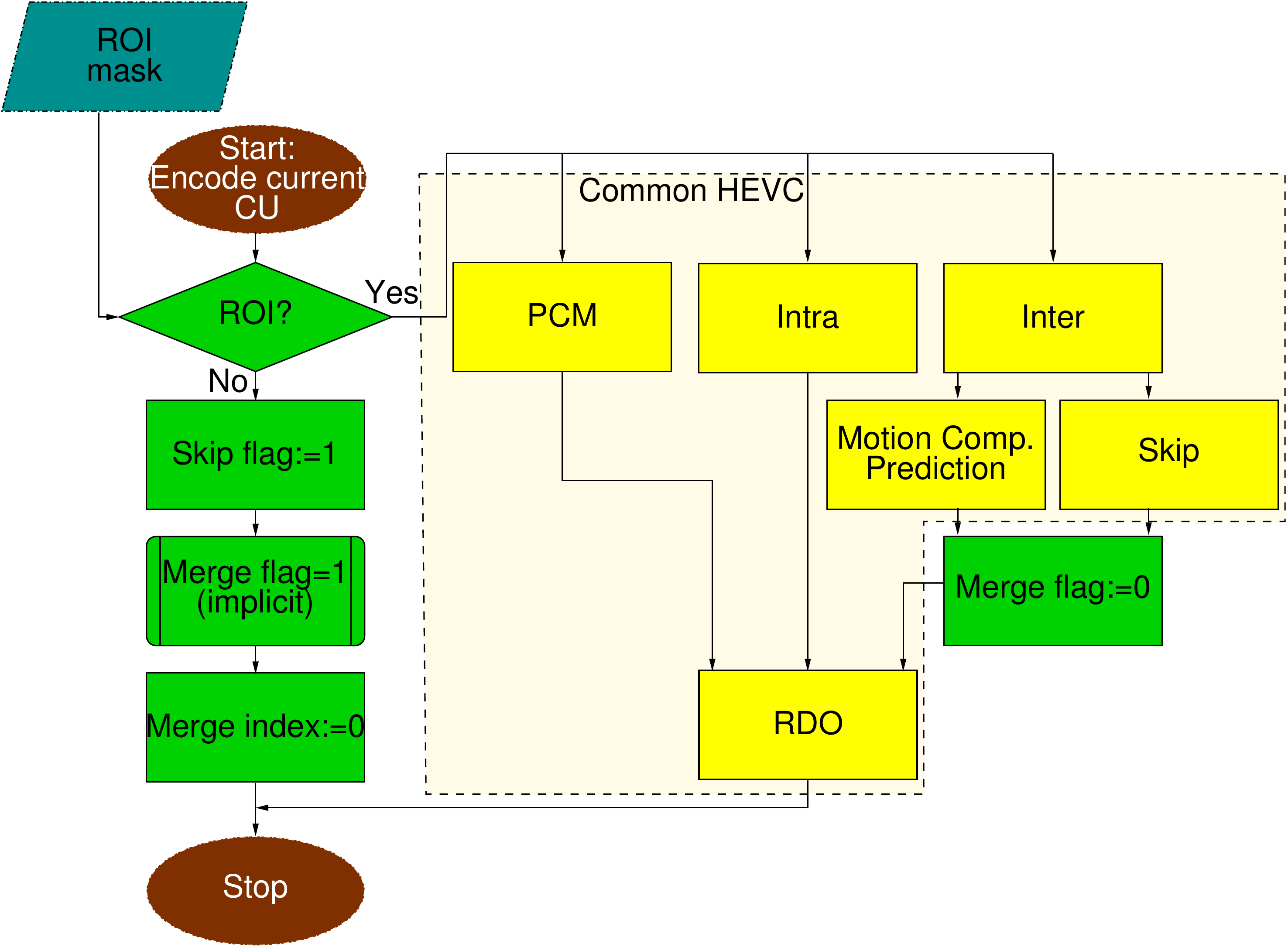} } %
  \caption{Flowchart of the \sc{HEVC}-skip coding system. Yellow: common \sc{HM}, green: proposed modifications, blue/top left: externally provided \sc{ROI} mask, brown/ellipses: start/stop.
  \vspace*{-1.15\baselineskip}
  }
  \label{fig:flowchart_hevc-skip}
\end{figure}

\section{Proposed Video Encoder Implementation}
\label{ssec:hevcCodingBackendImplementation}

To incorporate the increased coding performance of \sc{HEVC} compared to \sc{AVC} \cite{coding_efficiency_vgl_versch_videocodierstandards_und_hevc_2012_short} we transfer an external skip mode control similar to the \sc{AVC} implementation (``\sc{AVC}-skip'') into \sc{HEVC} (``\sc{HEVC}-skip'') and replace the video encoder in the \sc{ROI} coding system (\figref{fig:roi-control-scheme}) \cite{mesh_roi-coding_journal_pcs2013_followup_meuel_munderl_reso}.
We distinguish two cases again: \sc{ROI} and non-\sc{ROI}.
Since we are not interested in any content of non-\sc{ROI} \sc{CU}{}s as explained in the last section, we force to use \emph{skip mode} regardless of any \emph{Rate-Distortion Optimization} (\sc{RDO}) assuming that there cannot be any other mode (\ie \sc{PCM}/intra/inter prediction) which saves more bits than skip mode.
By contrast, \emph{Coding Units} (\sc{CU}{}) containing \sc{ROI} are encoded as usual by \sc{HM}.
Since the skip mode in \sc{HEVC} implies the \emph{merge mode} as mandatory, allowing the inheritance of motion vectors from spatially or temporally neighboring prediction units \cite{overview_of_the_high_efficiency_video_coding_standard_hevc_sullivan_ohm_han_wiegand_short}, 
the merge mode has to be controlled as well. %
It has two syntax elements: the binary \emph{merge flag} and an integer \emph{merge index} indicating the rate-distortion optimized best motion predictor from a candidate list for the current \sc{CU}.
The merge flag only has to be transmitted for non-skip modes whereas the merge index has to be transmitted for every skipped (and merged) block.
To retain standard compliance of the bit stream while minimizing the coding cost for a skipped \sc{CU} we force the merge mode a constant value (zero) for non-\sc{ROI} blocks (\figref{fig:flowchart_hevc-skip}, left/green column) in order to reduce the bit rate after \sc{CABAC} encoding.
For \sc{ROI} blocks we perform a normal \sc{RDO} with the only difference that for skip mode the merge mode is disabled completely (merge flag set to zero) to prevent prediction with a non-\sc{ROI} \sc{CU}.
The flow diagram is depicted in \figref{fig:flowchart_hevc-skip}.
Since the \sc{ROI} mask relies on $16 \times 16$ pel macroblocks in \sc{AVC}-skip, we propose to restrict the reference software \sc{HM} to \emph{Coding Tree Units} (\sc{CTU}{}s, formerly \emph{Largest Coding Units}, \sc{LCU}{}s) of $16 \times 16$ pels and a maximum partition depth of $2$ resulting in smallest \sc{CU}{}s of $4 \times 4$ pels.
Coding results for bigger \sc{CTU}{}s and higher partition depths (down to $4 \times 4$-\sc{CU}{}s) are additionally presented for \sc{HEVC}/\sc{HEVC}-skip for comparison. %

\vspace*{-\baselineskip}

\section{Experimental Results}
\label{sec:experiments}

\begin{table*}[ht]
  \caption{Coding gains (negative numbers) of proposed \sc{HEVC}-based over \sc{AVC}-based \sc{ROI} coding system compared to the reference (\emph{Ref.}) as marked in the table column by column. \sc{AVC} and \sc{HEVC} coding data rates without \sc{ROI} coding are additionally given (\sc{LD} configuration based \cite{hm_common_test_conditions_bossen_2013} with modified \sc{CTU} size/maximum partition depth). 
  Coding results for bigger block sizes are given for \sc{HEVC}. \emph{\sc{NS}}: \emph{non-splitted} \sc{CTU}{}s: \sc{CTU}{}s containing any \sc{ROI} are always \emph{entirely} encoded in a non-skip mode, \emph{\sc{HM}-subskip}: only those (small) \sc{CU}{}s containing \sc{ROI} are encoded in a non-skip mode, the remaining \sc{CU}{}s containing non-\sc{ROI} are encoded in the highly efficient skip mode.
  }
  \label{tab:coding_gains}
  \begin{scriptsize}
    \hspace{-1em}
      \begin{tabular}{ l c | r r r | r r r | r r r | r r r}
                    & & \multicolumn{3}{|c|}{\textbf{350\,m sequence}}
                      & \multicolumn{3}{|c|}{\textbf{500\,m sequence}}
                      & \multicolumn{3}{|c|}{\textbf{1000\,m sequence}} 
                      & \multicolumn{3}{|c}{\textbf{1500\,m sequence}} \\
                    & & \multicolumn{3}{|c|}{\textbf{43\,pel/m, 821\,frames}}
                    & \multicolumn{3}{|c|}{\textbf{30\,pel/m, 1121\,frames}}
                    & \multicolumn{3}{|c|}{\textbf{15\,pel/m, 1166\,frames}} 
                    & \multicolumn{3}{|c}{\textbf{10\,pel/m, 1571\,frames}} \\
                    & & \multicolumn{3}{|c|}{\textbf{\sc{PSNR} $\mathbf{\approx}$ 38.9\,dB}}
                    & \multicolumn{3}{|c|}{\textbf{\sc{PSNR} $\mathbf{\approx}$ 37.2\,dB}}
                    & \multicolumn{3}{|c|}{\textbf{\sc{PSNR} $\mathbf{\approx}$ 37.7\,dB}} 
                    & \multicolumn{3}{|c}{\textbf{\sc{PSNR} $\mathbf{\approx}$ 37.6\,dB}} \\
        Coder         & CTU          & \mccl{Data rate}               & \mcc{Diff.}          & \mcc{Diff.}         & \mccl{Data rate}               & \mcc{Diff.}          & \mcc{Diff.}          & \mccl{Data rate}               & \mcc{Diff.}          & \mcc{Diff.}          & \mccl{Data rate}               & \mcc{Diff.}          & \mcc{Diff.}          \\
        & in \unit{pel} & \mcc{in kbit/s}   & \mcc{in \%}         & \mccr{in \%}        & \mcc{in kbit/s}   & \mcc{in \%}         & \mccr{in \%}    & \mcc{in kbit/s}   & \mcc{in \%}         & \mccr{in \%}   & \mcc{in kbit/s}   & \mcc{in \%}         & \mcc{in \%} \\
        \hline
        AVC (x264)              & $16 \times 16$  &  9287         & \mcc{\emph{Ref.}} & \mccr{---}      &   11491                         & \mcc{\emph{Ref.}}   & \mccr{---}  &  9420                         & \mcc{\emph{Ref.}}   & \mccr{---}   & 13560                         & \mcc{\emph{Ref.}}   & \mcc{---}      \\
        HEVC (LD)               & $16 \times 16$  &  6489         & $-30.1$           & \mccr{---}      & 8973                          & $-21.9$             & \mccr{---}  & 7243                        & $-23.1$             & \mccr{---}    & 11942                          & $-11.9$             & \mcc{---}      \\
        HEVC (LD)               & $64 \times 64$  &  5568         & $-40.0$           & \mccr{---}       & 7947                          & $-30.8$             & \mccr{---}   & 5849                         & $-37.9$             & \mccr{---}  & 11901                          & $-12.2$             & \mcc{---}      \\
        \hline
        
        \hline
        AVC-skip                & $16 \times 16$  &   943         & $-89.8$           & \mccr{\emph{Ref.}}&  1423                         &  $-87.6$                   & \mccr{\emph{Ref.}}     &   1153                         &  $-87.8$             & \mccr{\emph{Ref.}} & 967                        & $-92.9$      & \mcc{\emph{Ref.}}    \\
        \hline
        HEVC-skip           & $16 \times 16$  &  \textbf{634} & $\mathbf{-93.2}$  & $\mathbf{-32.8}$  & \textbf{938}                         & $\mathbf{-91.8}$             & $\mathbf{-34.1}$  &  \textbf{797}                         & $\mathbf{-91.5}$             & $\mathbf{-30.9}$ & \textbf{664}                      & $\mathbf{-95.1}$             & $\mathbf{-31.3}$  \\ 
        HEVC-skip (NS)  & $32 \times 32$  &  659          & $-92.9$           & $-30.1$  & 987                          & $-91.4$             & $-30.6$ &  872                          & $-24.4$             & $-90.7$  & 836                          & $-93.8$             & $-13.6$ \\ 
        HEVC-skip (NS)  & $64 \times 64$  &  829          & $-91.1$           & $-12.1$  & 1338                           & $-88.4$             & $+42.6$     & 1172                           & $-87.6$             & $+1.7$   & 1335                           & $-90.2$             & $+38.1$   \\
        HEVC-subskip  & $64 \times 64$  & 559           & $-94.0$           & $-40.7$ &  853                          & $-92.6$             & $-40.1$    & 743                          & $-92.1$             & $-35.6$   & 616                            & $-95.5$             & $-36.2$   \\ 
        \hline \hline
      \end{tabular}
  \end{scriptsize}
\end{table*}

\newcommand{\sfigwidth}{0.47\linewidth} %
\begin{figure*}
  \centering
  \subfloat[\emph{350\,m sequence}, 43\,pel/m. \label{sfig:example_350m}]
  {
    \includegraphics[width=\sfigwidth]{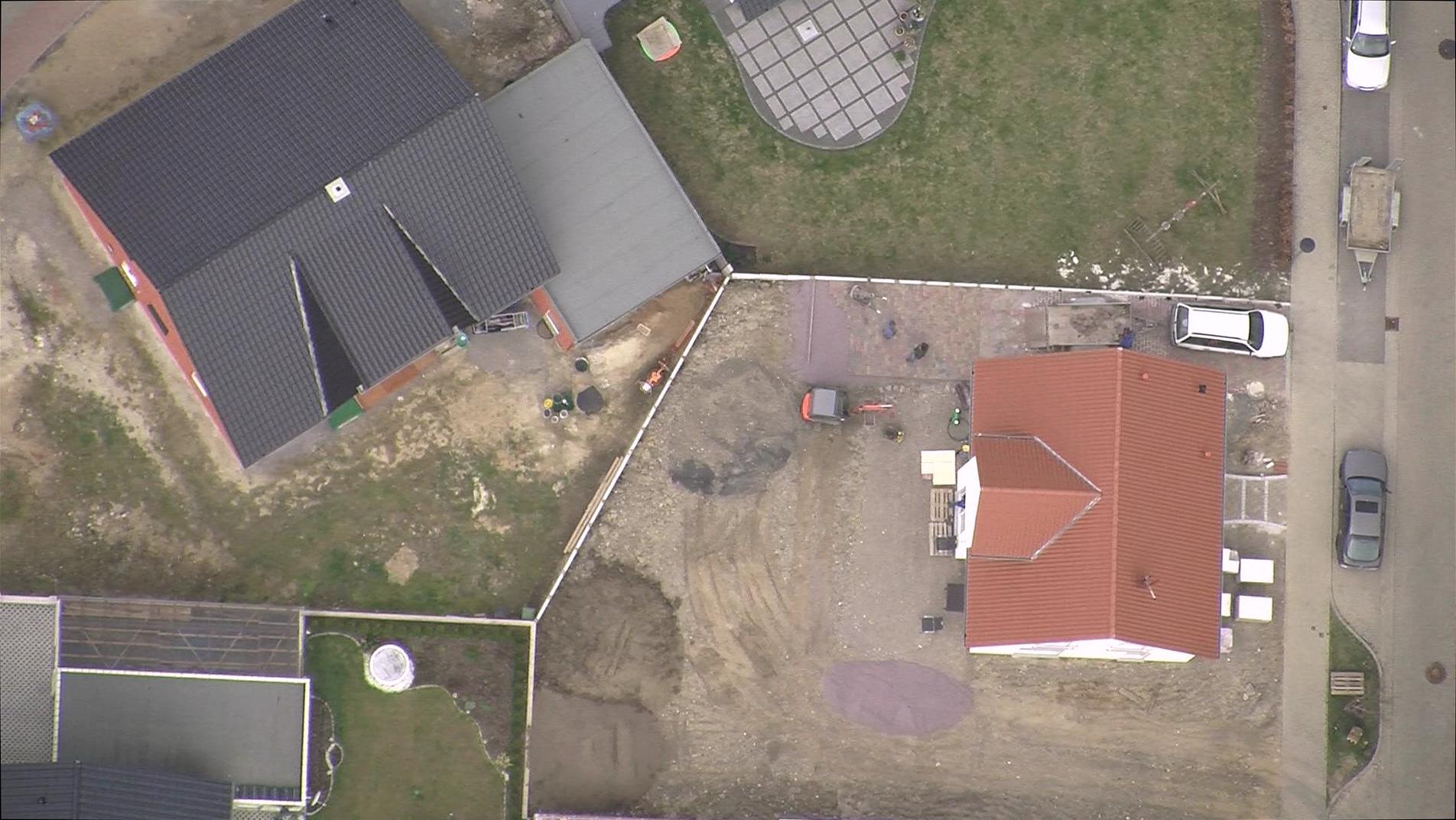}
  }
  \quad
  \subfloat[\emph{500\,m sequence}, 30\,pel/m. \label{sfig:example_500m}]
  {
    \includegraphics[width=\sfigwidth]{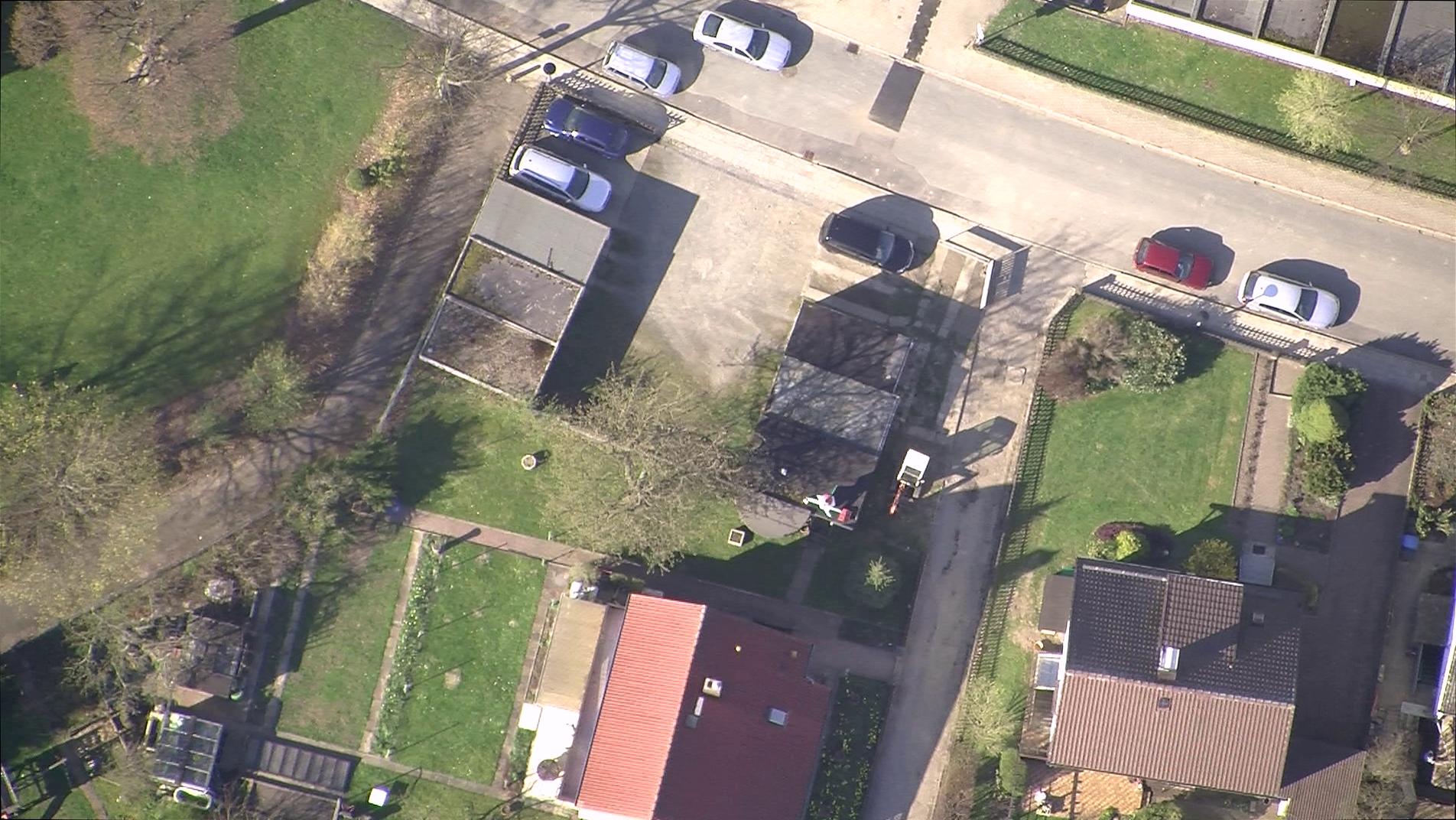}
  }
  \\
  \subfloat[\emph{1000\,m sequence}, 15\,pel/m. \label{sfig:example_1000m}]
  {
    \includegraphics[width=\sfigwidth]{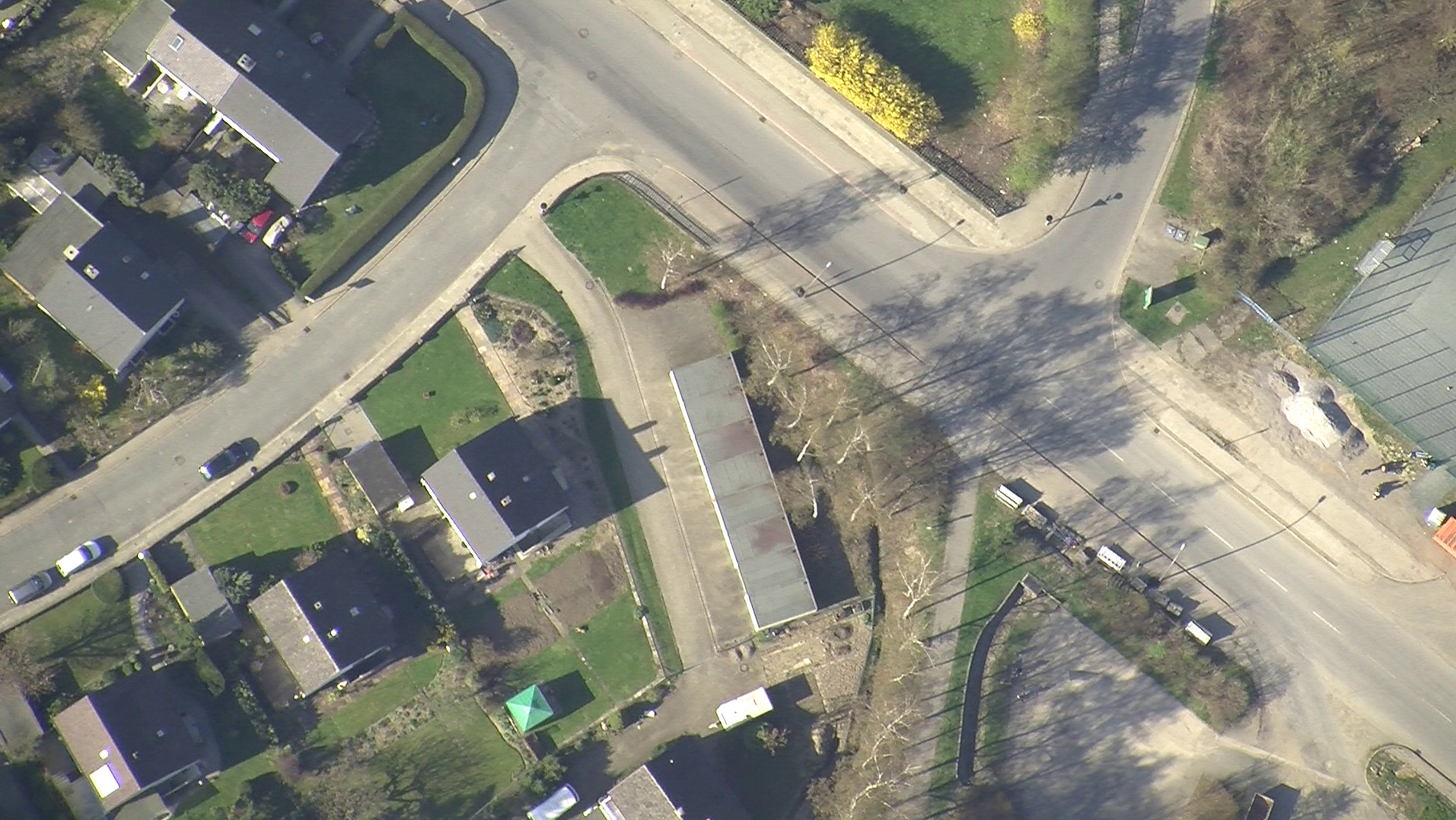}
  }
  \quad
  \subfloat[\emph{1500\,m sequence}, 10\,pel/m. \label{sfig:example_1500m}]
  {
    \includegraphics[width=\sfigwidth]{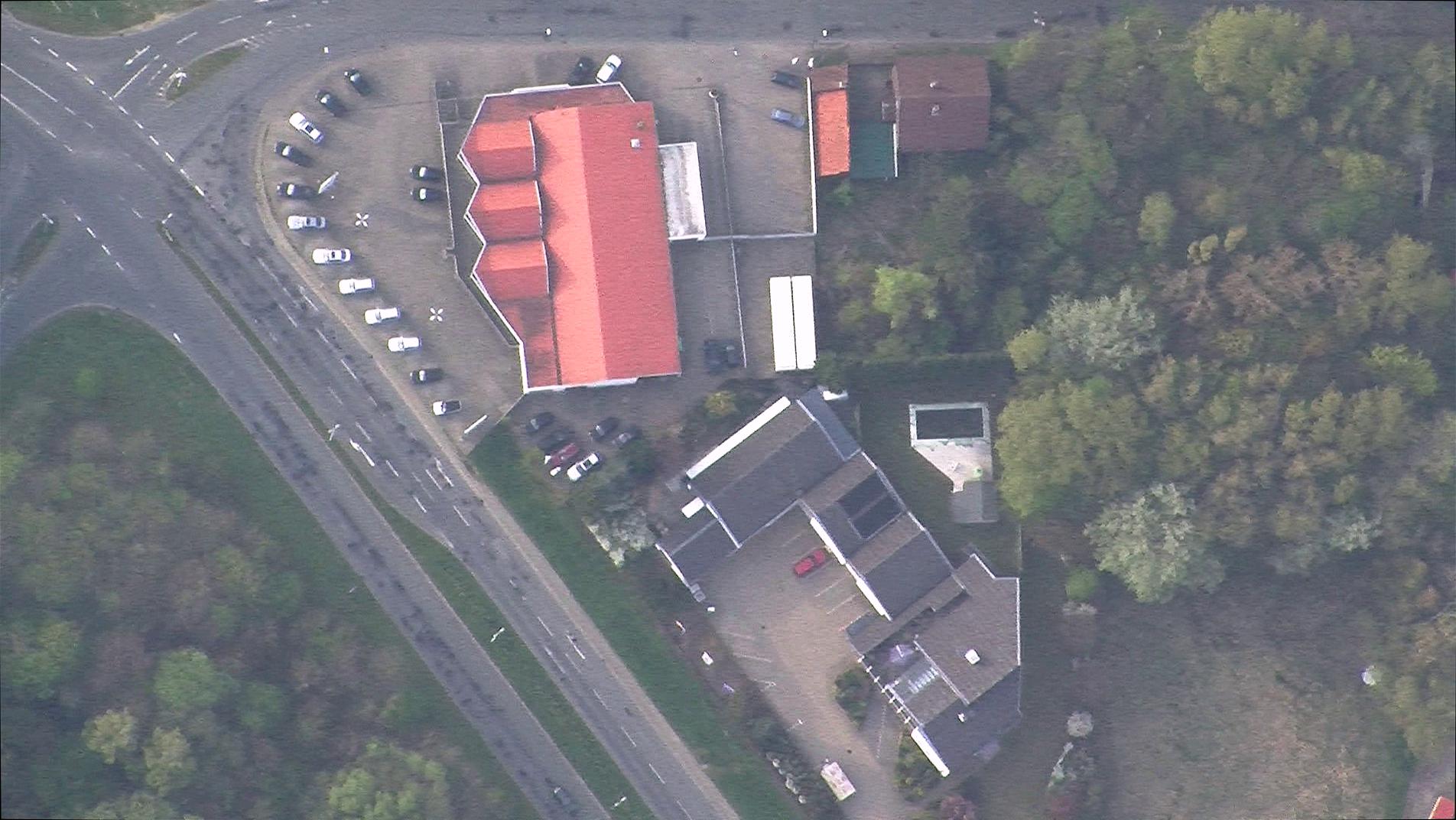}
  }
  \caption{Example frames from the test sequences with flight height and ground resolution \cite{tnt_aerial_video_testset_tavt}.
    \label{fig:exampleframes_planar}
  \vspace*{-\baselineskip}
  }
\end{figure*}

\begin{figure}[t]
  \centering
  \scalebox{0.95}{
    \begin{tikzpicture}
    \node [inner sep=0pt,outer sep=0pt] {
      \includegraphics[width=0.5\textwidth] {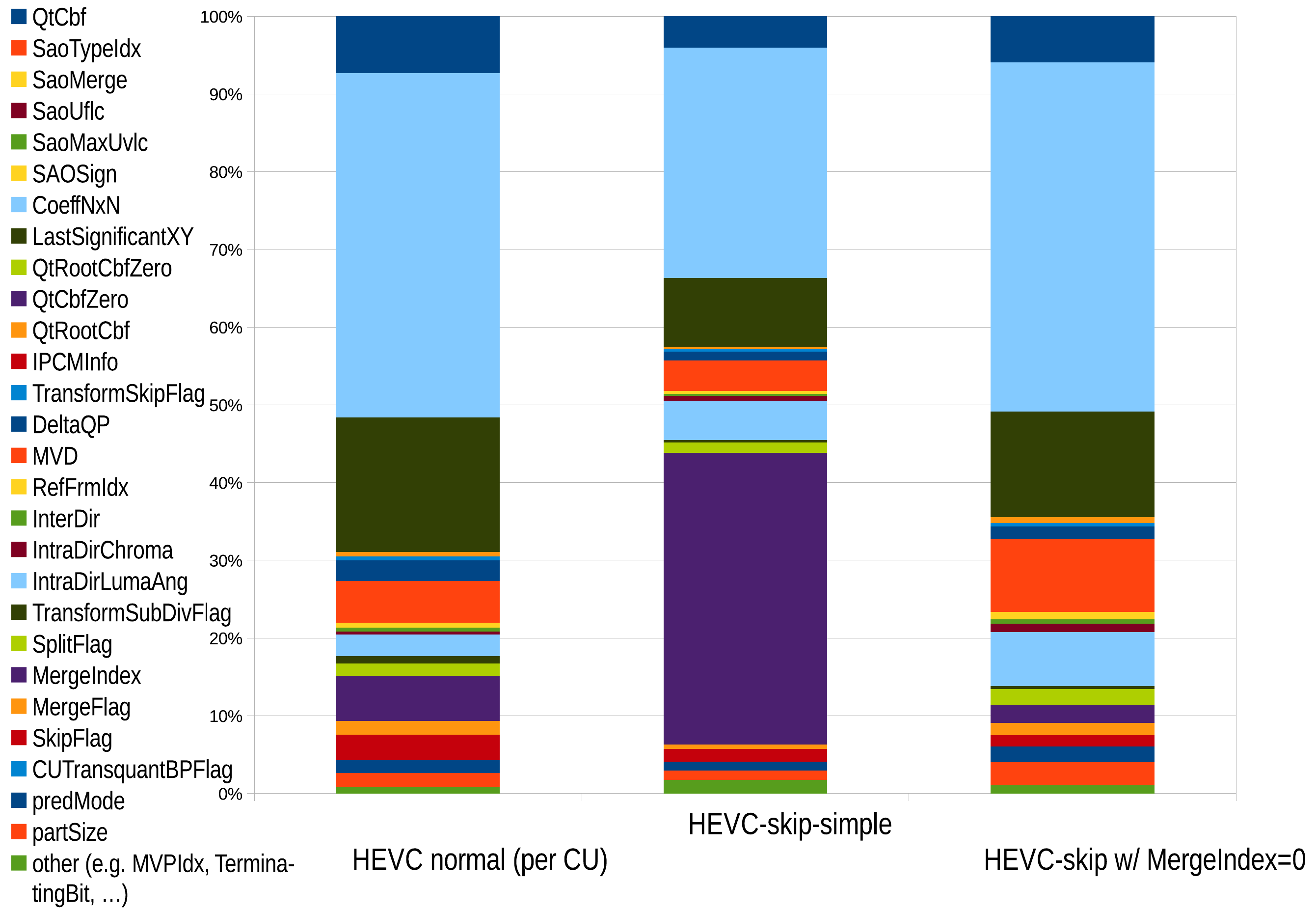}
    };
    \draw[draw=green,very thick]  (-1.62,-1.67) ellipse (0.8cm and 0.22cm);
    \draw[draw=red,very thick]  (0.6,-1.01) ellipse (0.8cm and 1.15cm);
    \draw[draw=green,very thick]  (2.9,-1.78) ellipse (0.8cm and 0.12cm);
    \end{tikzpicture}
  }
  \caption{Comparison of relative data rate consumption of syntax elements in the \sc{HEVC} bit streams for normal \sc{HEVC}/\sc{HEVC}-skip-simple/\sc{HEVC}-skip w/ merge index handling (per \sc{CU}).
  \vspace{-1.5\baselineskip}
  }
  \label{fig:hevc_skip_pp_comparison}
\end{figure}

The same detection results like for the \sc{AVC}-skip encoder are provided as input for the \sc{HEVC}-skip encoder (``ROI mask'' in \figref{fig:flowchart_hevc-skip}) and the coding performance of the \sc{AVC}-skip (modified \emph{x264}, v0.78 \cite{x264_software}) and the \sc{HEVC}-skip video encoder (based on \sc{HM} 10.0 \cite{hm_encoder_software_description_l1002_short}, modifications according to \secref{ssec:hevcCodingBackendImplementation}, \emph{low delay} (\sc{LD}) based configuration \cite{hm_common_test_conditions_bossen_2013} with modified \sc{CTU} size/maximum partition depth) are compared directly. %
As a reference also the unmodified \sc{HEVC} encoder (\sc{HM} 10.0) is compared to the unmodified x264 (v0.78) \sc{AVC} encoder \tabref{tab:coding_gains}.
We used 4 self-recorded \sc{HDTV} ($1920 \times 1080$, \unit[30]{fps}, consumer camera with global shutter) aerial video sequences from suburban areas from different flight heights (\unit[350]{m}, \unit[500, 1000, 1500]{m}, \figref{fig:exampleframes_planar}) resulting in corresponding ground resolutions of \unitfrac[43, 30, 15, 10]{pel}{m} (\emph{\sc{TNT} Aerial Video Testset} (\sc{TAVT}), \cite{tnt_aerial_video_testset_tavt}). %
The test sequences have different characteristics such as varying amount of \sc{ROI}{}s due to various sizes of \sc{ROI-NA} and changing numbers of moving objects like pedestrians and cars.
For the highest flight altitude the video is relatively noisy due to growing dusk.

For a reliable data rate comparison we measured the (luminance) \sc{PSNR} only for \sc{ROI} blocks for all ``skip-implementations'' assuming that the background quality stays constant due to the postprocessing (including \sc{GMC}).
The \sc{QP} for the \sc{HEVC} implementations were altered to match the bit rate of the \sc{AVC}-skip implementation with \sc{QP}$=25$ as close as possible.

Coding results for the different test sequences are provided in \tabref{tab:coding_gains}.
It is obvious that the average coding gain of \hevcskipimprovementmeaninteger (and also \hevcskipimprovementbigblocksmeaninteger for \sc{CTU}{}s of size $64 \times 64$) is lower than literature references \cite{coding_efficiency_vgl_versch_videocodierstandards_und_hevc_2012_short}, %
since only small parts of each frame (typically \unit[5--20]{\%}) are actually encoded in non-skip modes (all \sc{ROI} areas) and consequently are available for inter prediction.
Additionally the coding efficiency is limited by forcing smaller block sizes than allowed by the standard \cite{hevc_v1_standard}.
Coding results for bigger block sizes ($32 \times 32$ and $64 \times 64$) are also presented in \tabref{tab:coding_gains} for comparison. \emph{\sc{NS-CTU}} or \emph{non-splitted \sc{CTU}{}s} means that \sc{CTU}{}s containing any \sc{ROI-CU} are not splitted but entirely encoded in non-skip modes, leading to unnecessary encoded (non-\sc{ROI}) areas.
As a consequence the coding performance is decreased compared to \emph{\sc{HM}-subskip} (\sc{CTU}{}s containing \sc{ROI} may be further splitted in skipped/non-skipped \sc{CU}{}s).
Consequently for \emph{\sc{NS-CTU}} implementations the smallest (external enforced) skip block is equal to the \sc{CTU} size whereas it is $16 \times 16$ for the \emph{\sc{HM}-subskip} implementation.
We also tested a predictive encoder configuration based on the \sc{HEVC} \emph{Random Access} (\sc{RA}) configuration with hierarchical coding which performs similar to the \sc{LD} configuration.   %
For an \emph{All Intra} (\sc{AI}) configuration the relative gain is fairly constant at approximately \unit[25]{\%} but of course at a notable higher total bit rate.
It is salient that the coding gain of \sc{HEVC}-skip ($16 \times 16$ \sc{CTU}{}s) over \sc{AVC}-skip is also constant (about \hevcskipimprovementmeaninteger, \tabref{tab:coding_gains}, bold numbers) whereas the gains of unmodified \sc{HEVC} over \sc{AVC} vary in a wide range from \unit[11.9--30.1]{\%}, which can be assumed as typical considering different sequence characteristics (\eg noise) \cite{hevc_performance_analysis_dec2012_vanne_viitanen_hamalainen_hallapuro} and the reduced \sc{CTU} size.
Whereas the coding gains of the unmodified \sc{HEVC} are up to \unit[30]{\%} for sequences containing very little noise (\eg in the \unit[350]{m} sequence) we only gain about \unit[12]{\%} for a noisy and highly textured sequence (\unit[1500]{m} sequence).
The \sc{ROI} areas mainly contain new content (\sc{NA} is located on the left side for the test frame from the \unit[350]{m} sequence, and on the left and top side for the \unit[1500]{m} sequence) which is predominantly intra coded anyway (\figref{fig:cutypeusage}, note also the high amount of intra coded blocks (red dots) in non-\sc{ROI} for the \unit[1500]{m} sequence in \figref{sfig:cutypeusage_1500m}).
Those \sc{ROI} areas consume a high amount of bits which can be seen in the bit distribution maps in
 \figref{fig:heatmap}, especially for the \unit[350]{m} sequence.
Blue colors within these ``heat maps'' correspond to low bit usage for an \sc{CTU} whereas red colors indicate high bit usage.
The gain of \sc{HEVC}-skip over \sc{AVC}-skip is much higher than the gain of \sc{HEVC} over \sc{AVC} for the \unit[1500]{m} sequence than for the \unit[350]{m} sequence.
In order to predict the coding efficiency gain of aerial video sequences, we analyze the sequence characteristics.
Therefore we define the cost of coding individual blocks.
With the \emph{\sc{ROI}-bit-ratio} $C$ \eqref{eq:roicostratio} and the \emph{\sc{ROI}-area-ratio} $A$ \eqref{eq:roiarearatio} we define the \emph{bit-distribution-ratio} $R$ according to \eqref{eq:bitdistributionratio}.
  \begin{equation}
  \label{eq:roicostratio}
  C = \frac{\text{\sc{ROI} bit cost}}{\text{total bit cost}} \hspace*{-0.42em}
  \end{equation}
  \begin{equation}
  \label{eq:roiarearatio}
  A = \frac{\text{\sc{ROI} area}}{\text{frame area}} \hspace*{-0.42em}
  \end{equation}
  \begin{equation}
  \label{eq:bitdistributionratio}
  R = \frac{C}{A} \hspace*{-0.35em}
  \end{equation}

The difference in relative coding gains from \sc{HEVC} over \sc{AVC} compared to \sc{HEVC}-skip over \sc{AVC}-skip depends on very diverse ratios of $R$ for different sequences meaning that the bit usage for \sc{ROI}{}s drastically differs from the corresponding \emph{areas} covered by those \sc{ROI}{}s.

\newcommand{\subfigurewidth}{0.48\linewidth}
\begin{figure}[tb]
  \centering
  \hspace*{-5pt}
  \subfloat[\emph{350\,m sequence, \sc{ROI} left}. \label{sfig:cutypeusage_350m}]
  {
    \includegraphics[width=\subfigurewidth]{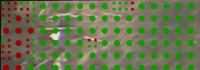}
  }
  \subfloat[\emph{1500\,m sequence}, \sc{ROI} left and top. \label{sfig:cutypeusage_1500m}]
  {
    \includegraphics[width=\subfigurewidth]{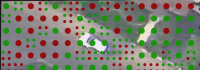}
  }
  \caption{Prediction modes of \sc{HEVC} (red dots: intra, green: inter, outtakes, \sc{ROI-NA} left in (a) and left/top in (b)).
    \label{fig:cutypeusage}
  \vspace{-1.5\baselineskip}
  }
\end{figure}
\newcommand{\subfigwidth}{\subfigurewidth} %
\begin{figure}[t]
  \centering
  \hspace*{-5pt}
  \subfloat[\emph{350\,m sequence}, $R=3$. \label{sfig:heatmap_350m}]
  {
    \includegraphics[width=\subfigwidth]{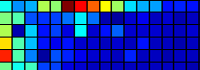}
  }
  \subfloat[\emph{1500\,m sequence}, $R=1.5$. \label{sfig:heatmap_1500m}]
  {
    \includegraphics[width=\subfigwidth]{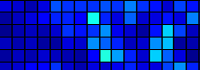}
  }
  \caption{Bit usage of example frames (``heat map'', outtakes).
    \label{fig:heatmap}
    \vspace*{-1.75\baselineskip}
  }
\end{figure}

If $R$ is $\approx 1$, the \sc{ROI} bit cost is proportional to the area covered by \sc{ROI} (\eg for the \unit[1500]{m} sequence with $R=1.5$, \figref{sfig:heatmap_1500m}).
If $R \gg 1$ the \sc{ROI} bit costs are unproportional high for \sc{ROI} areas, \ie a huge amount of bits consumed by one frame is used to encode only a small part of the frame (which is true for the other sequences with $3<R<4$). %
When such a frame is encoded with \sc{HEVC}-skip, the gain is much higher compared to the gain of \sc{HEVC} over \sc{AVC} like for this test set.
Consequently we can use the bit-distribution-ratio $R$ as an indicator for the \sc{HEVC}-skip coding gain relative to the unmodified \sc{HEVC} gain.
It is noteworthy that the encoding runtime decreases approximately linear with the number of blocks to be coded.
Thus, the encoding time of \sc{HEVC}-skip is decreased by \unit[80--95]{\%} compared to unmodified \sc{HM} for our test set. %
Despite additional processing time needed for global motion estimation and \sc{ROI} detection the entire detection \& coding system is much faster than \sc{HM}.

\section{Conclusions}
\label{sec:conclusion}

In this paper we propose to replace the \sc{AVC} video encoder by \sc{HEVC} in a \emph{Region of Interest} (\sc{ROI})-based coding system for aerial surveillance videos with a moving camera, \eg attached to an \sc{UAV}.
The coding system relies on an external control of the video encoder by \sc{ROI} detectors.
Only \sc{ROI} areas are regularly encoded whereas non-\sc{ROI} areas are forced to \emph{skip} mode.
We present an efficient mode control for \sc{HEVC} and can gain \hevcskipimprovementmeaninteger on average over an \sc{AVC}-skip implementation at similar coding block size and up to \hevcskipimprovementbigblocksmeaninteger for bigger coding block sizes (\sc{CTU} size of $64 \times 64$) which corresponds to coding data rates of \hevcskiprate at more than \unit[37]{dB} (\sc{ROI-PSNR}) for full \sc{HDTV} (\unit[30]{fps}) aerial surveillance video.
We provide a detailed analysis of spatial bit distribution of inter frames for the \sc{HEVC} encoder \sc{HM} and derive a bit-distribution-ratio as an indicator for the achievable coding gains of the proposed \sc{HEVC}-skip video encoder.
Results show highest relative gains of \sc{HEVC}-skip over \sc{AVC}-skip compared with \sc{HEVC} over \sc{AVC} for noisy and highly textured sequences.

\bibliographystyle{IEEEtran}
\bibliography{holgers_literaturdatenbank}

\end{document}